\documentclass[runningheads]{llncs}
\usepackage{graphicx}
\usepackage{amsmath,amssymb} 
\usepackage{color}
\usepackage[width=122mm,left=12mm,paperwidth=146mm,height=193mm,top=12mm,paperheight=217mm]{geometry}
\begin{document}
\pagestyle{headings}
\mainmatter
\def\ECCV18SubNumber{38}  

\title{Visual speech language models} 

\titlerunning{Visual speech language models}

\authorrunning{Bear H.L}


\author{Helen L Bear}
\institute{Queen Mary University of London, UK \& Technische Universit\"{a}t Mu\"{n}chen, Germany\\
			h.bear@qmul.ac.uk, dr.bear@tum.de}
\maketitle 

\section{Introduction}
\vspace{-.75em}
Language models (LM) are very powerful in lipreading systems (e.g. in \cite{bowden2013recent,6288999}). Language models built upon the ground truth utterances of datasets learn grammar and structure rules of words and sentences (the latter in the case of continuous speech). However, visual co-articulation effects in visual speech signals damage the performance of visual speech LM's as visually, people do not utter what the language model expects \cite{lieberman1963some}. These models are commonplace but while higher-order $N$-gram LM's may improve classification rates, the cost of this model is disproportionate to the common goal of developing more accurate classifiers. So we compare which unit would best optimize a lipreading (visual speech) LM to observe their limitations. As in \cite{bear2018alternative} we compare three units; visemes (visual speech units) \cite{lan2010improving}, phonemes (audible speech units), and words.

In the first two columns of Table~\ref{tab:sn_tests2} we list pairings of classifier units and language model units. For each pair we build a conventional lipreading system with the HTK toolkit \cite{htk} to classify Active Appearance model \cite{Matthews_Baker_2004} features extracted on 12 speakers from the RMAV audio-visual speech dataset\cite{bowden2013recent}. Phonemes are the International Phonetic Alphabet \cite{international1999handbook}, and our visemes are speaker-dependent visemes \cite{bear2017phoneme,bear2018comparing}. Word labels are from the RMAV ground truth. \textbf{Classifier units} are the labels used to identify individual classification models and \textbf{language units} make up the label scheme used for building the post-classification decoding LM. 

\section{Analysis} 
\begin{figure}[!hbt]
\centering
\begin{tabular}{lcr}
\includegraphics[width=0.33\textwidth]{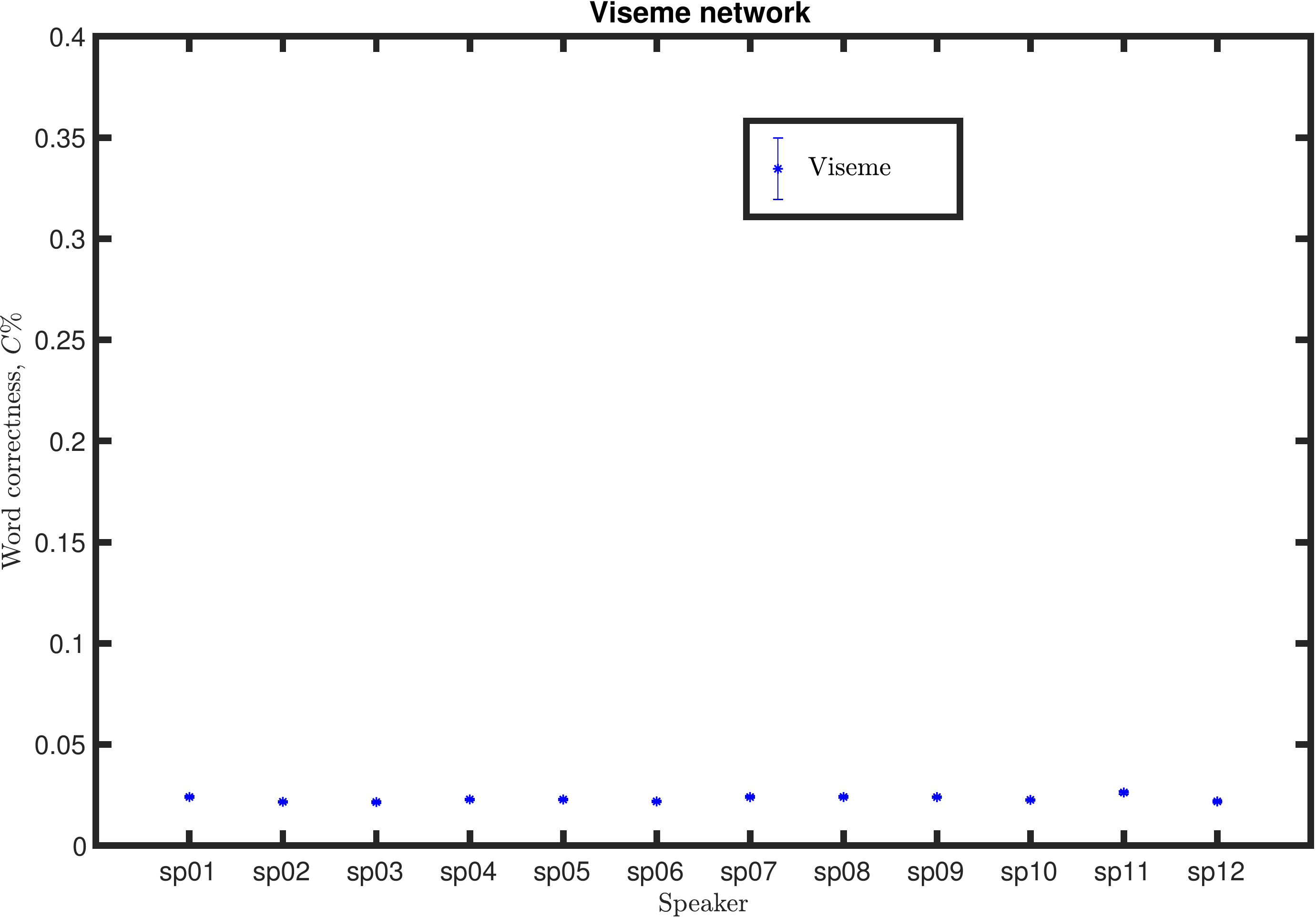} &
\includegraphics[width=0.33\textwidth]{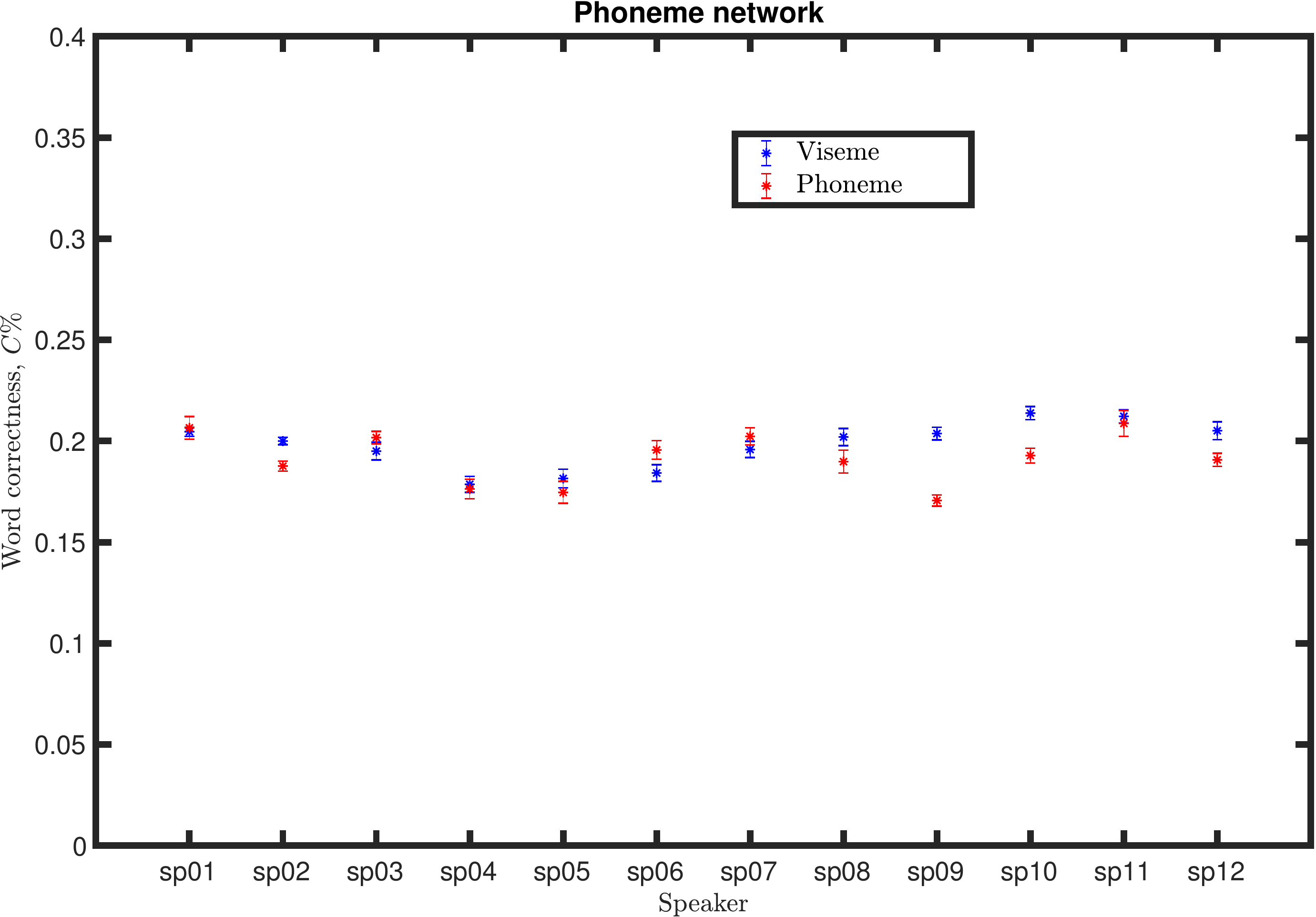} &
\includegraphics[width=0.33\textwidth]{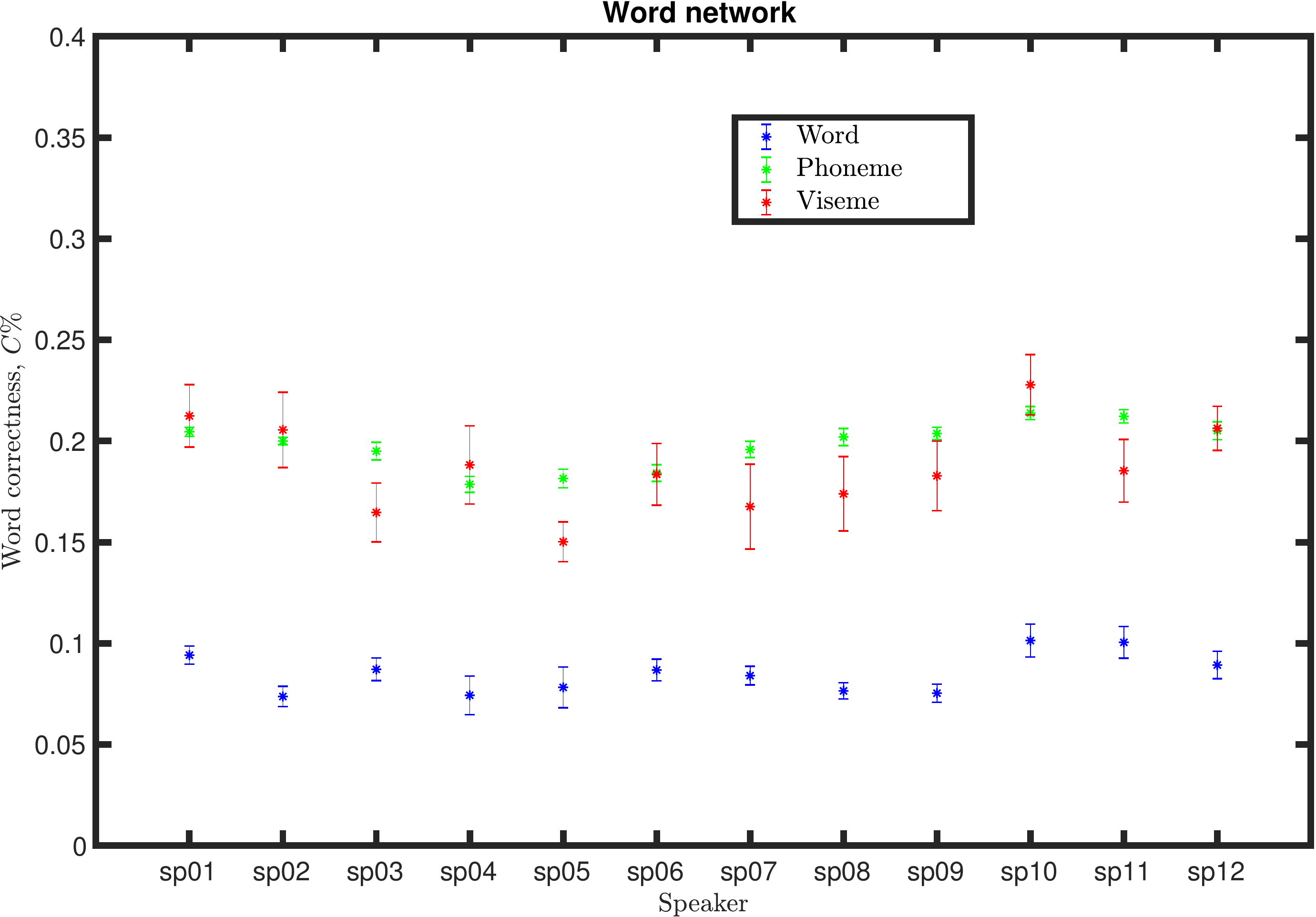} 
\end{tabular}
\caption{Effects of support network unit choice with each type of labeled HMM classifier units. Along the $x$-axis is each speaker, $y$-axis values are correctness, $C$.  Viseme LM is on the left, phoneme LM in the middle, and word LM on the right.}
\label{fig:sn_effects_talker}
\end{figure}

\vspace{-.75em}
Fig~\ref{fig:sn_effects_talker} shows word correctness (on the $y$-axis) for each speaker along the $x$-axis over three figures, one per LM unit. The viseme LM is on the left, phoneme LM middle, and word LM on the right. 
The viseme LM is the lowest correctness ($0.02\pm0.0063$). On the surface the idea of visemes classifiers is a good one because they take visual co-articulation into account to some extent. However as seen here, an LM of visemes is too complex due to the effect of homophemes \cite{thangthai2018comparing}.
The phoneme LM (Fig~\ref{fig:sn_effects_talker}, middle) is more exciting. For all speakers we see a statistically significant increase in $C_w$ compared to the viseme LM $C_w$ in Fig~\ref{fig:sn_effects_talker} left. Looking more closely between speakers, we see that for four speakers (2, 9, 10 and 12), the viseme classifiers outperform the phonemes, yet for all other speakers there is no significant difference. On average they are identical with an all-speaker mean $C_w$ of $0.19\pm0.0036$ (Table~\ref{tab:sn_tests2}). 

Lastly in Fig~\ref{fig:sn_effects_talker} (right) is $C_w$ of a word model paired with classifiers built on viseme, phoneme, and word units. Here word classifiers perform very poorly. We attribute this to insufficient training samples per class due to the extra number of classes in the word space ($>1000$ in RMAV) compared to the number of classes in the phoneme space ($49$) and so we do not recommend word-based classifiers without large volumes of visual speech data such as in \cite{chung2018learning,stafylakis2017deep}. Also in Fig~\ref{fig:sn_effects_talker} (bottom), are the phoneme and viseme classifiers (in green and red respectively) with a word LM. Here, for five of our twelve speakers (3, 5, 7, 8, and 11), the phoneme classifiers out-perform the visemes and for the other speakers there is no significant difference once a word LM is applied. This demonstrates that the strength of a good word network can help negate translations between acoustic and visual speech spaces. Fig~\ref{fig:sn_effects_talker} $C_w$ values and one standard error values are in Table~\ref{tab:sn_tests2}. This suggests phoneme units are most robust for visual speech language models but in practical terms this is not an easily intelligible output so words are preferred.
\begin{table}[!h]
\centering
\caption{All speaker mean $C_w$ for each pair of HMM and language model units.} 
\begin{tabular}{|l|l|r|r|}
\hline
Classifier units & Network units & $C_w$ & $1$se \\
\hline \hline
Viseme & Viseme & $0.02$ & $0.0063$		\\
Viseme & Phoneme & $0.19$ & $0.0036$ 		 \\
Viseme & Word & $0.09$ &  $0.0$		 \\
\hline
Phoneme & Phoneme & $0.19$ & 	$0.0036$ \\
Phoneme & Word & $0.20$ & 	$0.0043$	\\
\hline
Word & Word & $0.19$ & $0.0005$\\
\hline
\end{tabular}
\label{tab:sn_tests2}
\end{table}
\vspace*{-\baselineskip}
\section{Conclusion}
\vspace{-.75em}
For some speakers viseme classifiers with phoneme LMs are the better choice whereas others are easier to lipread with phoneme classifiers with a word LM. As experimenters we have no evidence to know which approach is best for a test speakers until after testing all unit options, so we recommend using either phoneme or word-based language network for future lipreading system development as these enable more interpretable prediction transcripts. Whilst lipreading LMs are powerful, they are not the solution to training machines to lipread \emph{all} speakers due to the great variation in speaker visual speech signals.

\bibliographystyle{splncs}
\bibliography{egbib}

\paragraph*{\footnotesize{This abstract is an extended extract from \textit{Alternative visual units for an optimized phoneme-based lipreading system}, Bear \& Harvey, Computer Speech and Language, 2018 (in review).}}

\end{document}